\documentclass{nature}
\usepackage{graphicx}

\title{Suppression of star formation in early-type galaxies by feedback from supermassive black holes}

\author{Kevin Schawinski$^{1}$,
Sadegh Khochfar$^{1}$, 
Sugata Kaviraj$^{1}$, 
Sukyoung K. Yi$^{2}$, 
Alessandro Boselli$^{3}$,
Tom Barlow$^{4}$,
Tim Conrow$^{4}$,
Karl Forster$^{4}$,
Peter G. Friedman$^{4}$,
D. Chris Martin$^{4}$,
Patrick Morrissey$^{4}$,
Susan Neff$^{5}$,
David Schiminovich$^{6}$,
Mark Seibert$^{4}$,
Todd Small$^{4}$,
Ted K. Wyder$^{4}$,
Luciana Bianchi$^{7}$,
Jose Donas$^{3}$,
Tim Heckman$^{7}$,
Young-Wook Lee$^{2}$,
Barry Madore$^{8}$,
Bruno Milliard$^{3}$,
R. Michael Rich$^{9}$ \&
Alex Szalay$^{7}$
}

\begin{document}

\maketitle

\begin{affiliations}
\item Department of Physics, University of Oxford, Oxford OX1 3RH, UK
\item Center for Space Astrophysics, Yonsei University, Seoul 120-749,
  Korea
\item Laboratoire d'Astrophysique de Marseille, 13376 Marseille Cedex
  12, France
\item California Institute of Technology, MC 405-47, Pasadena, CA
  91125
\item Laboratory for Astronomy and Solar Physics, NASA Goddard Space
  Flight Center, Greenbelt, MD 20771
\item Department of Astronomy, Columbia University, MC 5246, New York,
  NY 10027
\item Department of Physics and Astronomy, Johns Hopkins University,
  Baltimore, MD 21218
\item Observatories of the Carnegie Institution of Washington,
  813 Santa Barbara St., Pasadena, CA 91101
\item Department of Physics and Astronomy, UCLA, Los Angeles, CA 90095-1562
\end{affiliations}

\begin{abstract}
Detailed high resolution observations of the innermost
central parts of nearby galaxies have revealed
 the presence of supermassive black holes\cite{2000ApJ...539L..13G,
2001ApJ...547..140M, 2002ApJ...574..740T, 2005SSRv..116..523F}. These
 black holes may interact with their host galaxies by means of
 'feedback' in the form of energy and material jets; this feedback
 affects the evolution of the host and gives rise to the observed
 relations between the black hole and the host \cite{1998A&A...331L...1S}. 
Here we report observations of the ultraviolet emissions of massive
early-type galaxies. We derive an empirical relation for a critical
black-hole mass (as a function of velocity dispersion) above which the
outflows from these black holes suppress star formation in their hosts
by heating and expelling all available cold gas. Supermassive black
holes are negligible in mass compared to their hosts but nevertheless
seem to play a critical role in the star formation history of galaxies.
\end{abstract}

The near-UV (NUV) detector of the Galaxy Evolution Explorer
satellite \textit{GALEX}\cite{2005ApJ...619L...7M} covers a range
in wavelength between $1771$ and $2831\AA$ and is extremely sensitive to
young stellar populations. With it, we can detect small mass
fractions of 1-3\% of young stars formed within the last billion
years\cite{2005ApJ...619L.111Y, 2006astro.ph..1029K}. This high
sensitivity allows us to trace ongoing residual star formation in
present day early-type galaxies (ETGs). We select a volume-limited
sample of visually inspected ETGs from the Sloan Digital Sky
Survey\cite{2000AJ....120.1579Y} that have been observed by GALEX. We
remove active galactic nucleus (AGN) candidates from our sample
to avoid confusion between the observed UV emissions from the AGN
and from young stars \cite{2006astro.ph..1036S}.

We use $NUV-r$ colour as a probe of small amounts of recent star
formation (RSF) and find that the fraction of ETGs that show signs
of RSF is strongly correlated with the stellar velocity dispersion
$\sigma$, (hereafter "RSF-$\sigma$ relation"),
but not with the luminosity of the galaxy (see Figure 1). 
The correlation with $\sigma$ cannot be explained by a
systematic trend of $\sigma$ with internal extinction, because such a
trend would be exact opposite\cite{2005AJ....129.2138L} to that observed.

We use a semi-analytic model of galaxy evolution in the
$\Lambda$CDM paradigm to interpret these results\cite{wf91, spr01,
2005MNRAS.359.1379K} (for model details see
Supplements). Our simulation includes all the systematic processes
that affect galaxy evolution and takes into account the different
merger and star formation histories of galaxies\cite{2005astro.ph..9375K}. When we run this
simulation and investigate the predicted amount of recent star
formation in low redshift ETGs, we find that all galaxies
underwent enough star formation within the last billion years to
be detected by the extremely sensitive GALEX NUV detector. Massive
galaxies continued to gain cold gas as fuel for star formation
from mergers with satellites and
through the cooling of hot gas present in those galaxies.

This overproduction of stars in massive galaxies is a known
problem in models of galaxy evolution. In order to prevent star formation, a
shutdown mechanism is required: cold gas must be heated up and
expelled to prevent it from turning into stars. Feedback from
supernovae (SNe) was once a prime suspect, but 
it turned out to be too low to be effective in massive galaxies which have deep
potential wells\cite{1986ApJ...303...39D}. All the systematics
implicit in the $\Lambda$CDM paradigm, including feedback from SNe
have already been included in our simulation, so we turn to
feedback from active galactic nuclei (AGN) powered by supermassive
black holes (SMBH) as the  remaining most likely mechanism sufficiently
powerful to shut down star formation in massive galaxies to
reproduce the results of our observations\cite{
2005Natur.433..604D, 2006MNRAS.365...11C, 2005ApJ...630..705H}.

Energetic outflows caused by gas falling onto the supermassive
black holes thought to reside at the centres of galaxies can
self-regulate the growth of the black hole and the star formation
in the galaxy by heating the available gas and blowing it
out\cite{1998A&A...331L...1S}. Numerical simulations have shown
that the energy input from a black hole in such an AGN phase is
capable of doing this\cite{2005MNRAS.361..776S}. As SNe feedback is
incapable of producing the quiescent galaxies we observe with GALEX,
we conclude that AGN feedback is the most likely process by which the
available gas is heated and expelled in massive ETGs.

In addition to this, the nature of the observed correlation with sigma
(see Figure 1) allows us
to infer more about the way in which AGN feedback regulates galaxy
evolution. The masses of black holes correlate strongly with various
properties of the host galaxies and the relation thought to have
the smallest intrinsic scatter is that between $M_{\bullet}$ and
the velocity dispersion $\sigma$ as a proxy for the depth of the
gravitational potential\cite{2000ApJ...539L..13G,
  2002ApJ...574..740T, 2005SSRv..116..523F}. By
considering the maximum strength of an active galactic nucleus
given in terms of the Eddington luminosity $L_{edd} \propto
M_{\bullet}$ and expressing the gravitational potential depth by
the stellar velocity dispersion $\sigma$, it is possible to derive
a minimum black hole mass for which the outflow is capable of
depleting all gas from the potential\cite{1998A&A...331L...1S}. According to our
observations, massive ETGs which have large values of $\sigma$,
are on average less likely to have undergone recent star formation
compared to less massive ETGs: the "RSF-$\sigma$ relation".  
Thus, for a given velocity
dispersion,  we assume that there is a \textit{critical supermassive black hole
mass} above which the feedback powered by the black hole is
sufficiently powerful to heat up and/or expel all gas in the host
galaxy, thus preventing it from forming any stars even at the 1\%
level.

The observed intrinsic scatter in the $M_{\bullet}$$-$$\sigma$ relation gives 
 a range of black hole masses for a given value of $\sigma$.
Hence, in order to reproduce the decreasing fraction of RSF ETGs
with increasing $\sigma$, 
the $M_{\bullet,c}$$-$$\sigma$ relation should be less steep than
the $M_{\bullet}$$-$$\sigma$ relation itself. We can derive an
empirical $M_{\bullet, c}$$-$$\sigma$ relation from our observed trend 
assuming a slope and scatter for the $M_{\bullet}$$-$$\sigma$ relation. The
$M_{\bullet}-\sigma$ relation is usually expressed as a power law
of the form $M_{\bullet} = a_0 \left(\sigma/200\right)^{a_1}$, so
we also try to find such parameters $a_0$ and $a_1$ 
in the new $M_{\bullet,c}$$-$$\sigma$ relation that result in the
observed "RSF-$\sigma$ relation". We generate black
hole masses for the galaxies in our sample by putting their
velocity dispersions into the $M_{\bullet}-\sigma$ relation given
by Tremaine et al.\cite{2002ApJ...574..740T} and perturbing the black
hole masses
by the scatter of the relation; we do this ten times for each
galaxy (Figure 2 (a)). 
Then we vary $a_0$ and $a_1$ until we find the best match 
to the observed RSF-$\sigma$ relation. 
We find that with 95\% confidence
$M_{\bullet,c} = (1.05^{+0.07}_{-0.12}) \times 10^{8}
(\frac{\sigma}{200})^{3.2^{+0.5}_{-0.6}}$. 
As we are studying a volume
and magnitude-limited sample, the $M_{\bullet,c}$$-$$\sigma$ relation derived from
the Monte-Carlo simulation is only an approximation and we need to
perform full simulations, as they are complete in $\sigma$ as well as
in absolute magnitude, to derive the actual
$M_{\bullet,c}$$-$$\sigma$ relation.

We perform a more realisitc simulation using our semi-analytic 
models of galaxy evolution in order to test the plausibility of the 
hypothesis of a critical black hole mass controlling star formation via 
AGN feedback. This is a standard semi-analytic simulation in the
$\Lambda CDM$ paradigm that  
already includes various heating and cooling processes (e.g., SNe).
Our model does indeed reproduce the observed trend, but only
when AGN feedback is included, and
the best agreement with the observations (Figure 1) is achieved by
the relation $M_{\bullet,c}=1.26 \times 10^8 (\frac{\sigma}{200})^{3.65}$
(Figure 2b), which is in good agreement with our initial guess. 
The agreement in the lowest $\sigma-$bin is poorer
than in the others due to an increased scatter in our simulated
Faber-Jackson relation relative to the observed
one\cite{2003AJ....125.1866B}. The supermassive black holes grown in
our simulation have an intrinsic scatter of a factor of 1.5 in mass
and are in good agreement with the relation suggested by Tremaine et
al.\cite{2002ApJ...574..740T}

As our $M_{\bullet,c}-\sigma$ relation is less steep than the
$M_{\bullet}-\sigma$ relation we expect an increasing fraction of
star forming early-type galaxies with decreasing velocity
dispersion until eventually no AGN will be sufficiently powerful
to shut down all star formation. Boselli et
al.\cite{2005ApJ...629L..29B} find that the \textit{NUV-V}
(optical) and \textit{NUV-H} (near IR) colour-magnitude relations
for the early-type population of the Virgo cluster shows precisely
this kind of trend: low luminosity - and therefore less massive -
Virgo early-type galaxies are increasingly bluer and eventually
all lie below the RSF cut-off. We use stellar
models\cite{2003ApJ...582..202Y} to convert our RSF criterion to
the \textit{NUV-V} colours of the Virgo galaxies and find that
below $\sigma \sim 80 \mbox{kms}^{-1}$ all early-types are
classified as RSF. Our simulations agree with this cutoff very
well.

The possible implications of the $M_{\bullet,c}-\sigma$ relation
on galaxy formation and the growth of black holes can be
summarized by three main regions (see Figure 3): First, the low
$\sigma$-region, in which the majority of black holes are too
small to be able to halt star formation in their host galaxies by
feedback during their initial accretion phase. During this phase,
black holes grow mainly by accreting gas at the Eddington rate and
are only limited in their growth by the amount of available fuel.
Star formation on the other hand is regulated by the energy output
from supernovae into the interstellar medium and will be
proportional to the surface density of the
gas\cite{1998ApJ...498..541K}. This region extends up to $\sigma
\sim 80 \mbox{km s}^{-1}$ as suggested earlier. The next phase is
at $80 \mbox{kms}^{-1} < \sigma < 240 \mbox{kms}^{-1}$. Black
holes are now becoming on average massive enough to be able to
terminate star formation by their feedback and at the same time
limit their available fuel. The mass growth of the black holes is
still dominated by accretion but mergers between black holes start
playing an increasingly important role. Most host galaxies will be
identified as AGN or QSO when observed during the accretion phase
onto the black hole. We sub-divide this region into two: at the
lower end, a minority of galaxies will have critical black holes;
the black hole still grow mostly via
accretion. At the high end, most black holes are critical and so
most of the mass growth now occurs via mergers. The boundary
between these two is at the intersection between the
$M_{\bullet}-\sigma$ and the critical relation, i.e. $\sigma \sim
165 \mbox{kms}^{-1}$. Finally, for galaxies with $\sigma > 240
\mbox{kms}^{-1}$ the vast majority of black holes is massive
enough to halt all star formation and so star formation becomes
virtually prohibited in such massive galaxies. Both the galaxy and
the supermassive black holes now mainly grow via dry
mergers\cite{2003ApJ...597L.117K}. Further observations of nearby
galaxies hosting supermassive black holes in combination with star
formation estimates will help to shed more light into the scenario
we propose and will place a strong test for the existence of ``a
critical black hole mass'' in galaxies.


\begin{addendum}
 \item We would like to thank John Magorrian for discussions and comments.
GALEX (Galaxy Evolution Explorer) is a NASA Small Explorer, launched
in April 2003. We gratefully acknowledge NASA's support for construction,
operation, and science analysis for the GALEX mission, developed in
cooperation with the Centre National d'Etudes Spatiales
of France and the Korean Ministry of Science and Technology.
This work was supported by grant No. R01-2006-000-10716-0 from the Basic
Research Program of the Korea Science \& Engineering Foundation to S.K.Y. 
 \item[Competing Interests] The authors declare that they have no
competing financial interests.
 \item[Auhtor Information] K.S., Sadegh Khochfar, Sugata Kaviraj, 
S.K.Y. have performed the data
sampling, analysis, interpretation, model construction, and writing of
the manuscript. A.B. supplied the Virgo galaxy data, and the rest of 
the authors contributed to the data acquisition using the GALEX satellite.
Correspondence and requests for materials
should be addressed to S.K.Y. (email: yi@yonsei.ac.kr).
\end{addendum}

\newpage

\begin{figure*}
  \includegraphics[angle=90,width=0.9\textwidth]{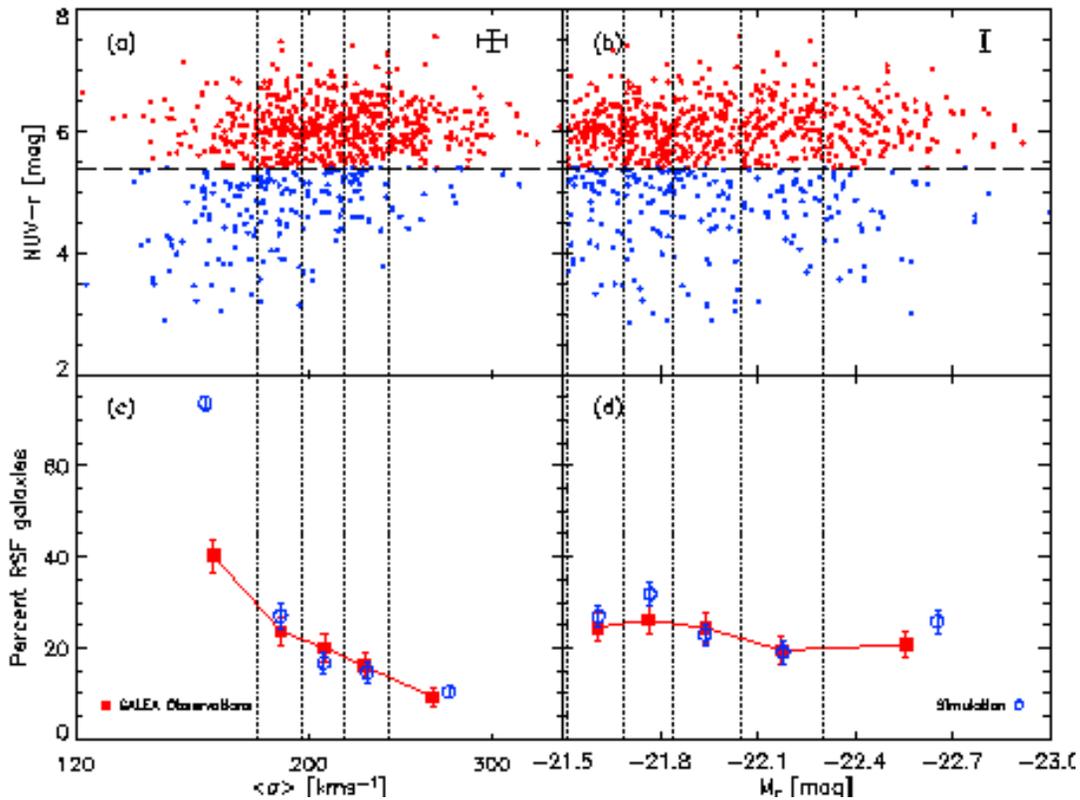}
  \caption{\textbf{The relationship between stellar velocity
      dispersion, luminosity and the fraction of star-forming early-type
      galaxies} 
      In panels (a) and (b), we show the observed UV colour-sigma and
      colour-magnitude relations. In the case of the colour-sigma
      relation, the fraction of blue early-type galaxies declines
      sharply as a function of sigma. In panels (c) and (d), we
      show the derived fractions of RSF (recent star formation)
      galaxies as a function of velocity dispersion and absolute
      magnitude. In red, we show the observed fractions, which
      include those early-type galaxies so faint in the UV that
      they are not detected by the \textit{GALEX} NUV detector.
      The blue points are the fractions of those galaxies in the
      model that correspond to having had between 1 and 3\% of
      stellar mass growth within the last 1 billion years. Error
      bars show Poisson errors. The vertical lines delineate
      the equal-number $\sigma$ bins. }
\end{figure*}

\newpage

\begin{figure*}
  \includegraphics[angle=90,width=0.9\textwidth]{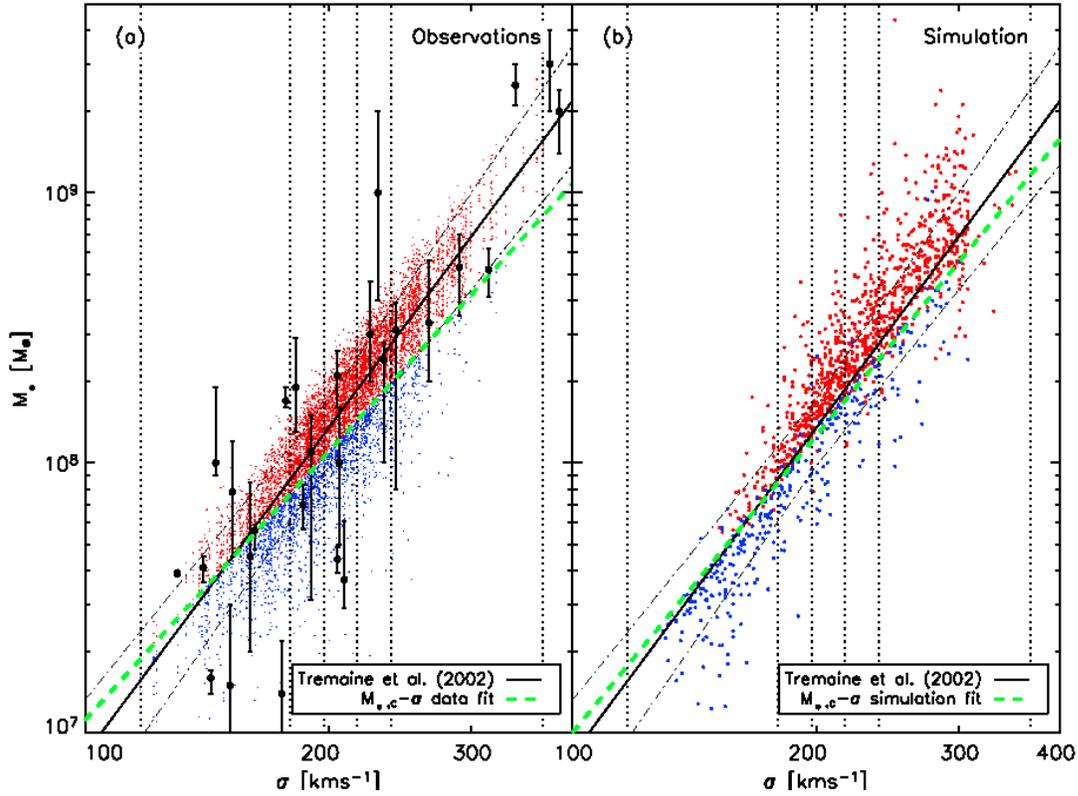}
  \caption{\textbf{The $M{_\bullet}-\sigma$ and the critical
    supermassive black hole mass-$\sigma$ relation} 
    (a) The observed $M{_\bullet}-\sigma$
    data points (black dots) with their errors taken from Tremaine et
    al.\cite{2002ApJ...574..740T} The solid line is the best fit to
    the $M{_\bullet}-\sigma$ relation, with the dash-dotted line
    showing the 1-$\sigma$ scatter. Based on this relation we perform
    a Monte-Carlo simulation for our galaxies (small dots).
    We colour these simulated
    galaxies depending on whether they lie above or below the critical relation
    in red (quiescent) and blue (RSF) that we derived by fitting the RSF
    fraction to the observed values (Figure 1). The vertical lines delineate
    the equal-number $\sigma$ bins of the data.
    (b) We show our semi-analytic galaxy models adopting
    $M_{\bullet,\,c}-\sigma$ relation shown in Figure 2(b),
    in comparison to the empirical $M_{\bullet}-\sigma$ relation.
    Red and blue dots are the model galaxies that did and did not
    have recent star formation in the last billion years.
    A new demarcation slope that effectively separates these galaxies
    has a $\sigma^{3.65}$ dependence (dashed green) which is close
    to the Monte-Carlo one in (a).}
\end{figure*}

\newpage

\begin{figure*}
  \includegraphics[angle=90,width=0.9\textwidth]{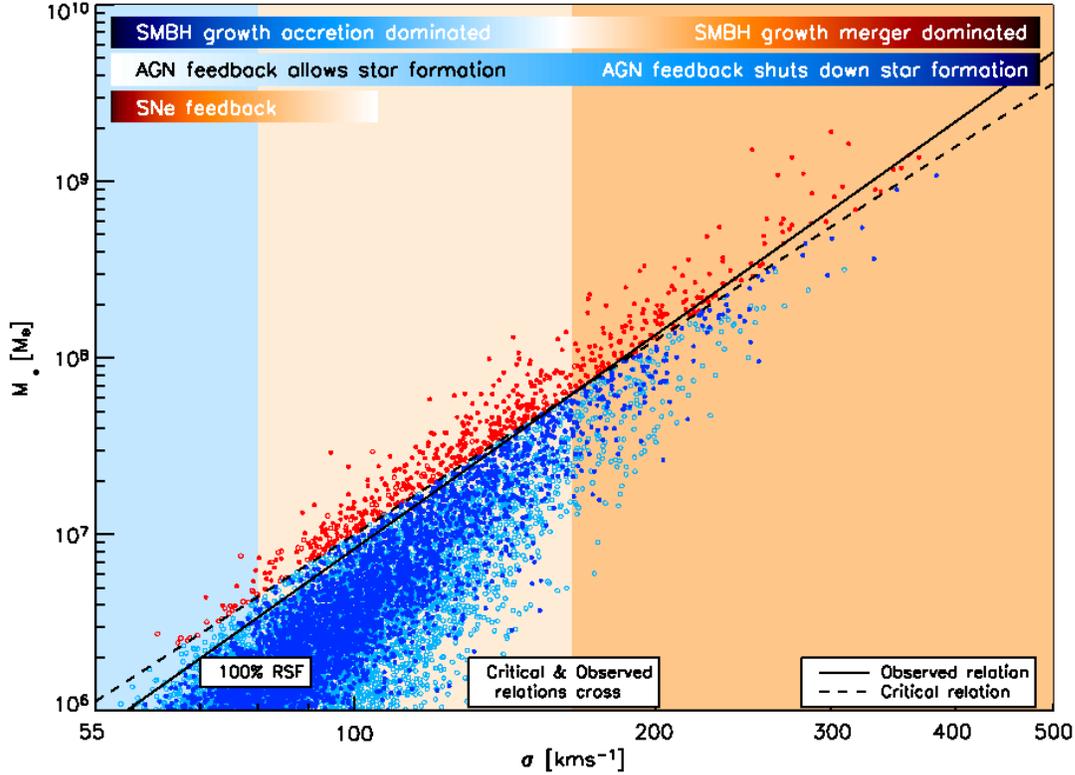}
  \caption{\textbf{A schematic view of how the critical supermassive
    black hole mass - velocity dispersion relation regulates galaxy evolution} 
    The black hole masses and velocity dispersions of our simulated
    galaxies down to $\sigma \sim 50 \mbox{kms}^{-1}$. The full points are
    early-type galaxies, while empty circles are late-type
    galaxies. Red and blue symbols denote quiescent and RSF galaxies,
    respectively. We show the observed $M_{\bullet}-\sigma$ relation
    (solid line) together with the best SAM critical relation (dashed). We
    indicate the various regimes by colour.  }
\end{figure*}

\newpage

\clearpage

\Large{\textbf{Supplements}}
\normalsize

\section{Galaxy Formation Modelling}
The main strategy behind the modelling approach we follow is to first calculate
 the collapse and merging history of individual dark matter halos, 
which is governed purely by  gravitational interactions, and secondly 
to calculate the more complex physics of the baryons inside 
these dark matter halos, including  e.g. radiative cooling of the gas, star formation 
and feedback from supernovae by simplified prescriptions on top of 
the dark matter evolution. Each of the dark matter halos consist of three 
main components which are distributed among individual galaxies inside  
them. A stellar, cold and hot gas component, where the latter  is
only attributed to {\it central} galaxies,  the most massive galaxies
inside individual halos.
In the following sections, we will describe briefly the recipes used to
 calculate these different components which 
are mainly based on recipes presented in Kauffmann et al.\cite{k99} (hereafter, K99) and 
Springel et al.\cite{spr01} (hereafter, S01), and we refer readers for more details 
on model implementations to their work and references therein.

Throughout this paper we use the following set of cosmological parameters:
$\Omega_0=0.3$, $\Omega_{\Lambda}=0.7$, $\Omega_b/\Omega_0=0.15$, 
$\sigma_8=0.9$ and $h=0.65$.

\subsection{Dark Matter Evolution}
We calculate the merging history of dark matter halos according to 
the prescription presented in Somerville et al.\cite{som99a}. This approach has 
been shown to produce merging histories and progenitor distributions in 
reasonable agreement with result from N-body simulations of cold dark matter 
structure formation in a cosmological context\cite{som00}. 
The merging history of dark matter halos is reconstructed by breaking 
 each halo up into progenitors above a limiting minimum progenitor 
mass $M_{min}$. This mass cut needs 
to be chosen carefully as it ensures that the right galaxy population and 
merging histories are produced within the model. Progenitor halos
 with masses below $M_{min}$ are declared as {\it accretion} events and 
their histories are not followed further back in time. 
Progenitors labelled as accretion events should ideally not host any 
significant galaxies in them and be composed mainly of primordial hot gas 
at the progenitor halo's  virial temperature. 
To achieve a 
good compromise between accuracy and computational time, we use
$M_{min}=10^{10}$ $M_{\odot}$, which ensures that the results presented here 
are unaffected by numerical resolution effects.

\subsection{Baryonic Physics}
As mentioned above, once the merging history of the dark matter 
component has been calculated, it is
 possible to follow the evolution of the baryonic content in these halos 
forward in time. We assume each halo  consists of three components: 
hot gas, cold gas and stars, where the latter two components can be distributed 
 among individual galaxies inside a single  dark matter halo. The stellar components 
of each galaxy are further divided into bulge and disc, to allow 
morphological classifications of model galaxies. In the following, we 
describe how the evolution of each component is calculated. 

\subsection{Gas Cooling \& Reionisation}\label{cool}
Each branch of the merger tree starts at a progenitor mass of $M_{min}$ and 
ends at a redshift of $z=0$. Initially, each halo is occupied by hot 
primordial gas which was captured in the potential well of the halo and 
shock-heated to its virial temperature 
$T_{vir}=35.9\left[V_c/(\mbox{km s}^{-1}) \right]^2$ K, where $V_c$ is the 
circular velocity of the halo\cite{k99,wf91}. Subsequently this hot 
gas component is allowed to radiatively cool and settles down into a
rotationally supported gas disc at the centre of the halo, which we identify 
as the central galaxy\cite{s77,wr78,wf91}.  
The rate at which hot gas cools down is estimated by calculating the 
cooling radius inside the halo using the cooling functions provided by 
\cite{sd93} and the prescription in S01. In the case of a merger 
between halos we assume that all of the hot gas present in the progenitors 
gets shock-heated to the virial temperature of the remnant halo, and this gas 
can cool down only onto the new central galaxy which is the central galaxy 
of the most massive progenitor halo. The central galaxy of the less massive 
halo will become a satellite galaxy orbiting inside the remnant halo. In this 
way, a halo can host multiple satellite galaxies, depending on the 
merging history of the halo, but will always only host one central galaxy onto 
which gas can cool. The cold gas content in  satellite galaxies  is given by 
the amount present when they first became satellite galaxies and does 
not increase, but decreases due to ongoing star formation and supernova 
feedback.

In the simplified picture adopted above, the amount of gas available to 
cool down is limited only by the universal baryon fraction 
$\Omega_b h^2=0.0224$\cite{sper03}. However, in the presence of a 
photoionising background the fraction of baryons captured in halos is reduced
\cite{ns97,g00} and we use the recipe of Somerville et al.\cite{som02}, 
which is based on a fitting formulae derived from hydrodynamical 
simulations\cite{g00}, to estimate the amount of baryons in each halo. 
For the epoch of  reionisation, we assume $z_{reion}=20$, which is in 
agreement with observations of  
the temperature-polarisation correlation of the cosmic microwave 
background\cite{ko03}.

\subsection{Star Formation in Discs and Supernova Feedback}\label{sf}
Once cooled gas has settled down in a disc, we allow for fragmentation and 
subsequent star formation according to a parameterised global 
Schmidt-Kennicutt law\cite{ken98} of the form 
$ \dot{M}_{*}=\alpha M_{cold}/t_{dyn,gal}$, where $\alpha$ is a free parameter
describing the efficiency of the conversion of cold gas into stars, and 
$t_{dyn,gal}$ is assumed to be the dynamical time of the galaxy and is 
approximated to be 0.1 times the dynamical time of the dark matter 
halo\cite{k99}.
As in K99 we allow star formation only in halos of $V_c < 350$ km/s to avoid 
excessively-massive central galaxies in clusters. 

Feedback from supernovae plays an important role in regulating star 
formation in small mass halos and in preventing excessively-massive 
satellite galaxies from forming. We implement feedback based on 
the prescription presented in K99 with
\begin{equation}
  \Delta M_{reheat}=\frac{4}{3} \epsilon \frac{\eta_{SN} E_{SN}}{V_{c}^{2}} 
  \Delta M_*. 
\end{equation}
Here we introduce a second free parameter $\epsilon$ which represents our 
ignorance about the efficiency with which the energy from supernovae 
reheats the cold gas.  The expected number of supernovae 
per solar mass of stars formed is given by $\eta_{SN}=5 \times 10^{-3}$,
taken as the value for the Scalo initial mass function\cite{sca86},  
 and $E_{SN}=10^{51}$ erg  is the energy output from each supernova. We take 
$V_c$ as the circular velocity of the halo in which the galaxy was  
last present as  a central galaxy.

\subsection{Galaxy Mergers}
We allow for mergers between galaxies residing in a single halo. As mentioned 
earlier, each halo is occupied by one central galaxy and a number of 
satellite galaxies depending on the past merging history of the halo. 
Whenever two halos merge, the galaxies inside  them are going to merge on a 
time-scale which we calculate by estimating the time it would take the 
satellite to reach the centre of the halo under the effects of 
dynamical friction. Satellites are assumed to merge only with 
central galaxies and we set up their orbits in the halo according to the 
prescription of K99, modified to use the Coulomb logarithm 
approximation of S01.

If the mass ratio between the two merging galaxies is $M_{gal,1}/M_{gal,2} 
\leq 3.5$ ($M_{gal,1} \geq M_{gal,2}$) we declare the event as a 
{\it major} merger and the remnant will be an elliptical galaxy and the 
stellar components and the gas will be treated according to the prescriptions 
below. In the case of {\it minor} merger $M_{gal,1}/M_{gal,2} > 3.5$ 
 the cold gas in the disc of the smaller 
progenitor is assumed to settle down in the gas disc of the remnant and its 
stars contribute to the bulge component of the remnant\cite{k99}.

\subsection{Formation of Ellipticals and Bulges}
It is suggested that major mergers will lead 
to the formation of elliptical galaxies\cite{tt72}. Indeed 
detailed numerical simulations in the 
last decade seem to support this hypothesis\cite{ba92,bn03}, and
we wil  assume in the following that major 
mergers disrupt the 
discs in progenitor galaxies as seen in various numerical simulations  
 and relax to a spheroidal 
distribution. During the merger, any cold gas in the discs of the progenitor 
galaxies is assumed to be funnelled into the centre of the remnant where it
ignites a starburst which transforms all of the cold gas into stars 
contributing to the spheroidal component\cite{k99,spr01}. The second 
assumption is certainly a 
simplification of what might happen since we neglect the possibility  that not all of the 
cold gas is funnelled to the centre but some fraction of it may for example 
settle down onto an extended disk 
 which continues to grow inside-out by fresh supply of gas from tidal 
tails\cite{bh91,mih96,ba01,b02}. 
The results of Barnes\cite{b02} indicate that $40\% - 80\%$ 
of the initial gas in the disc could end up in the central region of 
the remnant and be consumed 
in a starburst. The exact amount is somewhat dependent on the merger 
geometry and on the mass ratio of the merger. Unfortunately, a 
sufficiently large survey  
investigating the gas inflow to the centres of merger remnants is 
not available yet, so we use the simplified approach of assuming that
all cold gas gets used up in the central starburst.
This prescription for the fate of 
the cold gas  results in an overestimate of the spheroid masses and 
an underestimate 
of the secondary disc components in our model. This  is not very significant  
for massive ellipticals since they are mainly formed in relatively
non-dissipative mergers\cite{kb03}. 

\subsection{Formation and Growth of Super-Massive Black Holes}
We here follow the model introduced in Kauffmann \& Haehnelt\cite{2002MNRAS.332..529K} in which 
super-massive black holes get formed and fed during major mergers of glaxies. 
The assumption is to say that a fraction of the the cold gas available in the 
progenitor discs, that is funnelled into the centre of the remnant, 
will be accreted onto the 
black hole. Kaufmann \& Haehnelt introduced the following scaling law for 
the effectivness of this procces:
\begin{equation}
  \dot{M_{\bullet}}=\frac{f_{bh} M_{cold}}{1+(280 \mbox{km s}^{-1}/V_c)^2},
\end{equation}
where $M_{cold}$ is the amount of cold gas available in the progenitor disks,
 $V_c$ the circular velocity of the dark matter halo and $f_{bh}$ a free 
parameter. This scaling provides a good fit to the 
observed relation between the mass of the super-massive black hole and 
the velocity dispersion of a galaxy\cite{2002MNRAS.332..529K}. 
 Following Croton et al.\cite{2006MNRAS.365...11C} we here assume the velocity dispersion of a 
galaxy to be identical to the circular velocity of the dark halo the 
galaxy is embedded in. This however is not strictly true as e.g. in the case 
of an isothermal sphere $V_c/\sigma =\sqrt{2}$. Several authors have 
investigated the correlation between $V_c$ and $\sigma$\cite{2002ApJ...578...90F,2005ApJ...631..785P} and 
found different correlations. We here note that using a different 
scaling between $V_c$ and $\sigma$ in our models still produces the same 
$M_{\bullet}-\sigma-$relation if we adjust the free parameter $f_{bh}$ 
accordingly.

\subsection{Feedback from Super-Massive-Black Holes}
We here introduce a new prescription for the modelling of feedback from 
super-massive black holes, which is based on the results presented in this 
paper. According to our analyses a critical black hole mass exists for a given 
galaxy velocity dispersion $\sigma$, at which feedback is so strong that it 
reheats all the available cold gas and hence prevents further star formation. 
This critical black hole mass is derived from our empirical fit to the 
residual star formation fraction found in our data as 
\begin{equation}
  M_{\bullet,c}=1.26 \times 10^8 \left(\frac{\sigma}{200} \right)^{3.65}
\end{equation}
In galaxies with black holes more massive than the corresponding critical 
black hole mass we stop any cooling of gas and star formation.
We callibrate this relation to give the right fraction of RSF galaxies
as seen in Figure 1.

\subsection{Free Parameters}
We normalise our two model parameters for the star formation 
efficiency $\alpha$ and supernova feedback efficiency $\eta$  
by matching the $I-$band Tully-Fisher relation 
of Giovanelli et al.\cite{gio97} and requiring that spiral central galaxies of 
halos with circular velocity
$V_C=220$ km/s
have on average $10^{11}$ M$_{\odot}$ of stars and 
$10^9$ M$_{\odot}$ of cold gas\cite{som99b}. The third free parameter 
$f_{bh}$, which regulates the black hole growth is set to match the observed 
$M_{\bullet}-\sigma-$relation and is $f_{bh}=0.02$.

\section{The Virgo Cluster Data}

We derive the $NUV-V$ colour-$\sigma$ relation for the Virgo cluster
down to much smaller values of $\sigma$ than in our GALEX-SDSS sample
by combining the near-UV and optical photometry from Boselli et
al.\cite{2005ApJ...629L..29B} with velocity dispersion measurements
from the GOLDmine database\cite{2003A&A...400..451G}. 
The purpose of looking at the
Virgo cluster galaxies is to constrain the low end of the $\sigma$ and
so get an idea of the point at which all galaxies are RSF galaxies and
thus below critical. Since the optical photometry of the Virgo
galaxies in this sample are in V-band, we approximately convert the
RSF criterion of \textit{NUV-r}=5.4 to \textit{NUV-V}. We use simple
stellar populations of solar metallicity and three ages (3,6 and 9
Gyr) to derive the \textit{NUV-V} RSF criterion; these are \textit{NUV-V}=
5.16, 5.12 and 5.09 respectively. In all three cases, the point at
which no Virgo early-type galaxy is above this cutoff - i.e. all show
some signs of young populations - is at $\sigma \sim 80 kms^{-1}$. 

\section{The RSF criterion}

We use \textit{NUV-r} colour as a discriminant between those galaxies which
are quiescent and those which show signs of recent star
formation. Besides AGN, there is one further effect which can
mimic young stars in the near-UV. Extremely old
stellar populations can give rise to the UV upturn phenomenon in
some early-type galaxies\cite{1979ApJ...228...95C,
1988ApJ...328..440B}. We choose the \textit{NUV-r} colour of one of the
strongest UV upturn galaxies NGC 4552\cite{1988ApJ...328..440B}
to be the boundary between galaxies with no young component on the one hand and
those that are so blue they must have some young stars in
them. Although still quite limited, theoretical population synthesis
models which are supported by empirical data (Lee et
al.\cite{2005ApJ...619L.103L}) also suggest that the NUV-r of
passively evolving ETGs could not be much bluer than 5.4.
Using this criterion, we compute the fraction of galaxies that
must have experienced some star formation within the last billion
years. The ages and mass fractions of these young stellar
components are generally 300-500 Myr and 1-3\% by mass\cite{Kaviraj}. 
Due to the high sensitivity to dust extinction, these
fractions are a lower limit as cold gas and dust have been
detected in many early-type galaxies\cite{1989ApJS...70..329K,
2004ApJS..151..237T}. 
\textit{NUV-r} probes recent star formation for up to 1 Gyr and so
traces different time scales than those of AGN feedback. so there may
be a certain amount of time lag between the UV emissions and the
processes underlying feedback.

\section{Visual Inspection of Galaxy Morphology}

\begin{figure*}
  \includegraphics[angle=90,width=0.9\textwidth]{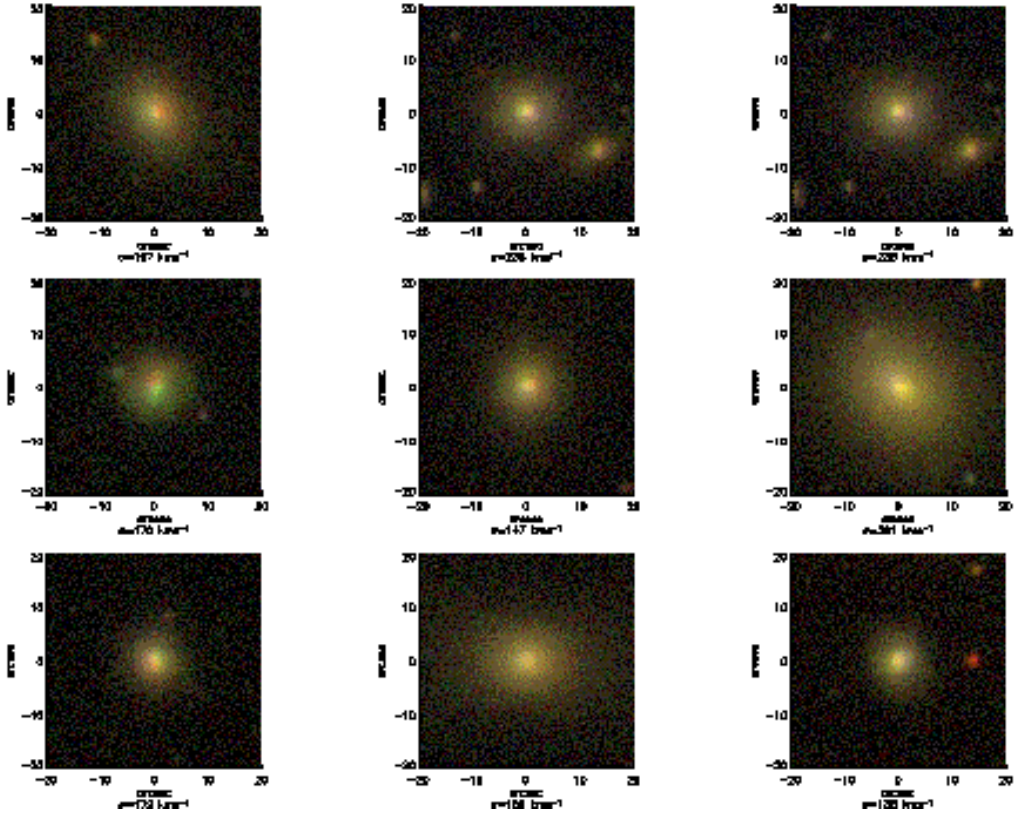}
  \caption{Example galaxies from our SDSS sample. The images are gri
  composites\cite{2004PASP..116..133L} and span a wide range in velocity dispersion.
    \label{vi_ex}}
\end{figure*}

The process and criteria of the visual inspection are described in
detail in Schawinski et al.\cite{2006astro.ph..1036S}. 
In summary, we find early-type galaxy (ETG) candidates from the SDSS
database using $frac\_Dev$ parameter, which is the weight of the 
de Vaucouleurs component in the two-component (de Vaucouleurs and
exponential disc) fits. We use $frac\_Dev>0.95$, which is a highly 
conservative criterion, hoping to exclude as many spiral interpolers
as possible. Despite this, still some spiral bulges remain in our sample.
Hence, we remove them via visual inspection on $gri$ bands\cite{2004PASP..116..133L}.
Since visual inspections are also subject to errors, we investigat 
out to what redshift and apparent magnitude visual inspection 
based on SDSS images is reliable by comparing them with one of 
the COMBO-17 fields overlapping SDSS. 
The COMBO-17 image is significantly deeper and has
much better seeing with 0.7" as opposed to 1.4" which
are typical for SDSS. By comparing our classification based on these
two different data sets, we concluded that visual inspection of
morphology using SDSS was possible to z $\sim$ 0.13 and R $<
17.31$. We settle on more conservative limits of z $<$ 0.1 and r $<$
16.8. The presence of late-type interlopers with star-forming disks or
spiral arms is therefore not a significant concern 
and does not give rise to the
correlations between $M_{r}$, $\sigma$ and NUV-r color. A selection of
sample images with their velocity dispersions are shown in Figure
\ref{vi_ex}.

\end{document}